\documentclass[11pt,a4paper]{article}

\usepackage[margin=1in]{geometry}
\usepackage{amsmath,amssymb}
\usepackage{bm}
\usepackage{graphicx}
\usepackage{tikz}
\usepackage{helvet}
\usepackage{authblk}
\usepackage[colorlinks=true, linkcolor=blue, citecolor=blue, urlcolor=blue]{hyperref}

\usetikzlibrary{positioning, fit, matrix, calc, arrows.meta}

\title{\textbf{AI Meets Turbulence Modeling: Complex Systems Perspective}\thanks{Accepted for publication in \textit{National Science Review}.}}

\author{Sharath S. Girimaji}
\affil{Department of Ocean Engineering, Texas A\&M University, College Station, TX 77843, USA \\ \textit{Corresponding author email:} \href{mailto:girimaji@tamu.edu}{girimaji@tamu.edu}}

\date{\today}

\begin{document}

\maketitle

\begin{abstract}
Despite widespread enthusiasm, applying artificial intelligence and machine learning (AI/ML) to turbulence modeling has yielded only modest progress, primarily due to a lack of situation-aware deployment. Data-driven Reynolds-averaged Navier--Stokes (ML-RANS) closures frequently fail to generalize across diverse flow regimes, while scale-resolving simulations (SRS) remain computationally prohibitive. As a result, ML-RANS is often misapplied outside its domain of validity, and ML-SRS is unnecessarily deployed on flows fully amenable to RANS. To guide optimal AI/ML deployment, we propose a turbulence taxonomy map parameterized by degrees of freedom ($Re$) and the degree of non-equilibrium ($SK/\epsilon$). By applying Weaver's well-recognized complexity framework, we classify turbulence into distinct physical regimes. This categorization identifies where local ML-RANS surrogates can generalize, clarifies where explicit scale resolution is physically mandatory, and highlights where physics-informed neural networks (PINNs) are most effective. Ultimately, this perspective establishes a physics-grounded framework for the flow-aware, successful deployment of AI/ML in turbulence.
\end{abstract}

\vspace{0.5em}

\vspace{1.5em}


Integrating artificial intelligence and machine learning (AI/ML) into computational fluid dynamics has been heralded as a paradigm shift \cite{Duraisamy2019ARFM}. Researchers leverage deep neural networks and symbolic regression to develop data-driven closures for the Reynolds-Averaged Navier-Stokes (RANS) equations. Yet, despite an unprecedented level of effort, a fundamental generalization crisis persists \cite{rumsey2022nasa}. Current ML-RANS models often interpolate accurately within specific training manifolds, only to fail, sometimes drastically, when extrapolated across different geometries or forcing conditions. 
This ML-RANS fragility stems not from algorithmic limitations or insufficient data, but from an ill-posed physical premise. The prevailing ML-RANS paradigm treats turbulence as a uniform (independent and identically distributed, or i.i.d.) dataset, implicitly assuming a generalizable mapping from local mean flow statistics to unresolved Reynolds stresses. To overcome this ML-RANS limitation, machine learning techniques have also been developed in the scale-resolving simulations (ML-SRS) context. Furthermore, novel approaches such as Physics-Informed Neural Networks (PINNs) are showing significant promise. However, there is currently a lack of clarity or \emph{a priori} guidance regarding the appropriate ML approach for a given flow application. 

As these data-driven methodologies become increasingly prevalent, more formal prescriptions must be developed and deployed. Toward this end, the primary goal of this study is to develop an appropriate, physics-aware framework to guide the judicious choice of ML techniques. By establishing conceptual physical boundaries based on fundamental flow regimes, this perspective seeks to ensure that ML strategies are applied precisely where they are physically valid and computationally optimal.

In an incompressible fluid ($\partial u_i/\partial x_i = 0$), turbulence evolution is governed by:
\begin{equation}
\frac{\partial u_i}{\partial t} + u_j \frac{\partial u_i}{\partial x_j} = -\frac{1}{\rho}\frac{\partial p}{\partial x_i} + \nu\frac{\partial^2 u_i}{\partial x_j\partial x_j}.
\label{eq:NS}
\end{equation}
Applying a filtering operator $u_i=\overline{u}_i+u_i'$ yields the resolved-scale evolution equation \cite{Germano}:
\begin{equation}
\frac{\partial \overline{u}_i}{\partial t} + \overline{u}_j \frac{\partial \overline{u}_i}{\partial x_j} = -\frac{1}{\rho}\frac{\partial \overline{p}}{\partial x_i} + \nu\frac{\partial^2\overline{u}_i}{\partial x_j\partial x_j} - \frac{\partial\tau_{ij}}{\partial x_j},
\label{eq:RANS}
\end{equation}
where $\tau_{ij}$ is the unclosed generalized second-moment tensor, which reduces to the Reynolds stress tensor in the ensemble-average limit \cite{Germano}. The central challenge is the constitutive closure modeling of the $\tau_{ij}$ tensor to accurately capture the impact of unresolved scales on resolved motion.

\section{Constitutive Relation}

Crucially, this constitutive relation should not, in general, be regarded as an instantaneous function of the local mean velocity or strain-rate field \cite{Lumley1970}. It is more properly viewed as a spatio-temporal \emph{functional}: the stress at a given location and time may depend on the historical trajectory of the flow throughout its spatial neighborhood \cite{Lumley1970},
\begin{equation}
    {\tau}_{ij}(\boldsymbol{x},t)
    =
    \mathcal{F}_{ij}\left[
        \overline{\boldsymbol{u}}(\boldsymbol{x}',t')
    \right], \quad \text{for } t' \le t.
    \label{eq:lumley_functional}
\end{equation}
This formulation is analogous to the distinction in continuum mechanics between a history-dependent constitutive law in a nonlinear viscoelastic medium and an instantaneous material function. An instantaneous closure,
\begin{equation}
    {\tau}_{ij}(\boldsymbol{x},t)
    =
    \mathcal{G}_{ij}\!\left(
        \overline{\boldsymbol{u}}(\boldsymbol{x},t)
    \right),
    \label{eq:instantaneous_closure}
\end{equation}
is therefore a restricted representation of the broader constitutive functional. In an equilibrium state—where historical memory is rapidly dissipated and absorbed into universal equilibrium scales—the functional $\mathcal{F}_{ij}$ collapses to an instantaneous relation $\mathcal{G}_{ij}$. In non-equilibrium turbulence, however, the flow retains long-time memory of its strain and convective history. Replacing the functional with a local function in these non-equilibrium regimes inevitably discards physical information essential to the closure.

The Mori--Zwanzig (MZ) projection-operator formalism provides the exact mathematical framework for decomposing these unresolved dynamics into local Markovian and history-dependent memory contributions. The exact MZ projection for the resolved flow evolution is:
\begin{equation}
\frac{d\overline{\boldsymbol{u}}}{dt} = \mathcal{M}(\overline{\boldsymbol{u}}) + \int_0^t \mathcal{K}(t-s)\,\overline{\boldsymbol{u}}(s)\,ds + \mathcal{F}(t),
\label{eq:MZ}
\end{equation}
where $d/dt$ is the substantial derivative following the resolved flow, $\mathcal{M}$ represents local Markovian dynamics, $\mathcal{K}$ is the memory kernel, and $\mathcal{F}$ is the orthogonal fluctuating force. 

Because unresolved dynamics enter Eq.~(\ref{eq:RANS}) strictly via the stress divergence field ($-\nabla \cdot \boldsymbol{\tau}$), equating unclosed forces implies that the subgrid stress tensor inherits an exact structural decomposition,
\begin{equation}
\tau_{ij} = \tau_{ij}^{\,M} + \tau_{ij}^{\,K} + \tau_{ij}^{\,O},
\label{eq:stress_decomp}
\end{equation}
where $\tau_{ij}^{\,M}$ is the local Markovian response of rapidly decorrelating fluctuations (the basis for algebraic eddy-viscosity closures), $\tau_{ij}^{\,K}$ represents the non-local history of long-lived coherent structures that violate local closure assumptions \cite{Girimaji2024NJP}, and $\tau_{ij}^{\,O}$ is the residual stochastic forcing typically neglected or absorbed in deterministic closures.

This Mori--Zwanzig decomposition reveals that turbulent complexity is dictated by the dynamic interplay between linear forcing ($\mathcal{L}$) and non-linear relaxation toward statistical equilibrium ($\mathcal{N}$). This competition is governed by two fundamental physical timescales: the mean deformation time $\tau_L = O(1/S)$ (where $S = \sqrt{2S_{ij}S_{ij}}$ is the resolved strain-rate magnitude), which dictates the rate of structural generation; and the non-linear turbulent relaxation time $\tau_N = K/\epsilon$ (where $K$ is the turbulent kinetic energy and $\epsilon$ is its dissipation rate), which dictates how rapidly the turbulence decorrelates and forgets its history.

The ratio of these timescales defines a dimensionless metric,
\begin{equation}
\frac{SK}{\epsilon} = \frac{\tau_N}{\tau_L},
\label{eq:strain_timescale}
\end{equation}
which acts as a turbulent Deborah number \cite{Lumley1970} and quantifies the kinematic competition between structural organization and stochastic relaxation \cite{Girimaji2024NJP}. When $SK/\epsilon \ll 1$, stochastic non-linear relaxation dominates ($\tau_N \ll \tau_L$); fluctuations decorrelate rapidly before coherent organization can form, collapsing the memory kernel into a local Markovian response ($\tau_{ij}^{\,M}$). Conversely, when $SK/\epsilon \gg 1$, linear deformation outpaces relaxation ($\tau_L \ll \tau_N$), allowing long-lived coherent structures to persist within the unresolved field. In this regime, the turbulence exhibits viscoelastic-type memory, producing a broad memory kernel that necessitates the non-local, history-dependent contribution $\tau_{ij}^{\,K}$. The parameter $SK/\epsilon$ thus serves as the primary kinematic metric dictating whether a flow regime admits local closures or requires non-local, history-aware modeling.

\section{Turbulence Through the Lens of Complexity}

Turbulent flows have traditionally been classified by geometry and Reynolds number ($Re$), which reflects the system's active degrees of freedom. The central premise of the present framework is to augment this classical view by incorporating the flow's degree of non-equilibrium—manifested as memory-laden structural organization—into the classification scheme.

In his foundational taxonomy of scientific problems, Weaver \cite{weaver1948science} argued that multi-variable physical systems must be categorized not merely by their number of interacting variables, but by their degree of internal organization. Weaver established that systems with vast degrees of freedom fall into two distinct regimes: \textit{disorganized complexity}, where variables interact stochastically and are readily described by ensemble or spatial averages (e.g., classical thermodynamics), and \textit{organized complexity}, where variables exhibit strongly correlated, non-random structural organization that resists simple statistical averaging.

Translating Weaver's taxonomy to fluid mechanics, we construct a classification scheme that maps the competition between local stochastic response and non-local structural memory onto two orthogonal physical axes:
\begin{itemize}
    \item \textbf{Phase-Space Dimensionality (Degrees of Freedom):} Parameterized by the Reynolds number ($Re$), this axis represents the sheer number of active scales in the flow, governed by non-linear cascade processes ($\mathcal{N}$).
    \item \textbf{Kinematic Non-Equilibrium (Degree of Organization):} Parameterized by the timescale ratio $SK/\epsilon$, this axis reflects the departure from equilibrium driven by flow-aware linear deformation ($\mathcal{L}$), dictating the extent of coherent structural organization.
\end{itemize}

\begin{figure}[t]
\centering
\begin{tikzpicture}[
    font=\sffamily,
    >=Stealth,
    scale=1.0
]
\colorlet{r1color}{gray!12}
\colorlet{r2color}{green!12}
\colorlet{r3color}{orange!16}
\colorlet{r4color}{red!14}

\filldraw[fill=r1color, draw=gray!50, thick] (0,0) rectangle (4.6,3.6);
\filldraw[fill=r2color, draw=gray!50, thick] (4.6,0) rectangle (9.2,3.6);
\filldraw[fill=r3color, draw=gray!50, thick] (0,3.6) rectangle (4.6,7.2);
\filldraw[fill=r4color, draw=gray!50, thick] (4.6,3.6) rectangle (9.2,7.2);

\draw[->, very thick] (-0.2,0) -- (9.8,0);
\draw[->, very thick] (0,-0.2) -- (0,7.8);

\node[font=\footnotesize\itshape, anchor=north east] at (9.6,-0.05) {increasing $\rightarrow$};
\node[font=\footnotesize\itshape, anchor=south east, rotate=90] at (-0.05,7.6) {increasing $\rightarrow$};

\node[align=center, font=\bfseries, anchor=north] at (4.6,-0.5)
{Phase-Space Dimensionality, $Re$\\[-0.5mm]
{\normalfont\small Active Degrees of Freedom ($\mathcal{N}$)}};

\node[align=center, rotate=90, font=\bfseries, anchor=south] at (-0.5,3.6)
{Kinematic Non-Equilibrium, $SK/\epsilon$\\[-0.5mm]
{\normalfont\small Degree of Organization / Memory ($\mathcal{L}$)}};

\node[align=center, text width=4.3cm] at (2.3,1.8) {
\textbf{Regime I: Simplicity}\\[1.5mm]
\textit{Low $Re$, Low $SK/\epsilon$}\\[1.5mm]
Laminar / Viscous\\[1mm]
\textbf{Analytically Predictable}
};

\node[align=center, text width=4.3cm] at (6.9,1.8) {
\textbf{Regime II: Disorganized}\\[-0.5mm]
\textbf{Complexity}\\[1.5mm]
\textit{High $Re$, Low $SK/\epsilon$}\\[1.5mm]
Stochastic Equilibrium\\[1mm]
\textbf{Learnable / RANS Valid}
};

\node[align=center, text width=4.3cm] at (2.3,5.4) {
\textbf{Regime III: Organized}\\[-0.5mm]
\textbf{Simplicity}\\[1.5mm]
\textit{Low $Re$, High $SK/\epsilon$}\\[1.5mm]
Deterministic Coherence\\[1mm]
\textbf{Compute $>$ Learn}
};

\node[align=center, text width=4.3cm] at (6.9,5.4) {
\textbf{Regime IV: Emergent}\\[-0.5mm]
\textbf{Complexity}\\[1.5mm]
\textit{High $Re$, High $SK/\epsilon$}\\[1.5mm]
History-Dependent Dynamics\\[1mm]
\textbf{Resolve (SRS Required)}
};

\end{tikzpicture}
\caption{\textbf{Turbulence complexity taxonomy.} Adapting Weaver's framework \cite{weaver1948science}, flow regimes are classified by phase-space dimensionality ($Re$, horizontal axis) and kinematic non-equilibrium ($SK/\epsilon$, vertical axis). Regime II represents disorganized complexity, where stochastic fluctuations are near equilibrium, rendering data-driven RANS surrogates highly learnable and transferable. Conversely, Regime IV represents \textit{emergent complexity}. In this regime, pronounced memory effects and structural organization cannot be modeled \textit{a priori}; they mathematically preclude local closure, strictly demanding explicit scale-resolving simulation (SRS) rather than statistical surrogacy.}
\label{fig:quadrant}
\end{figure}
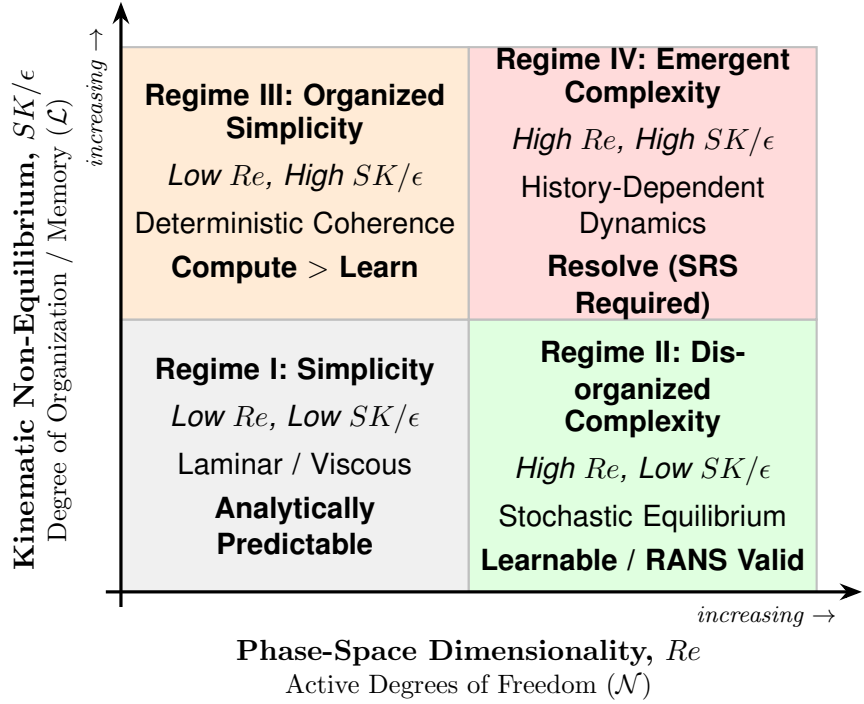

Projecting fluid motions onto this two-axis parameter space yields a four-regime complexity matrix (Fig.~\ref{fig:quadrant}) that directly maps Weaver's philosophical framework to the turbulence closure and ML generalization problem:

\noindent \textbf{Regime I: Simplicity (Low $Re$, Low $SK/\epsilon$):} Laminar or trivial flows governed entirely by predictable viscous diffusion. With few degrees of freedom, these flows are analytically tractable or computationally trivial, requiring no statistical modeling.
    
\noindent \textbf{Regime II: Disorganized Complexity (High $Re$, Low $SK/\epsilon$):} Fully developed, near-equilibrium stochastic turbulence. In this regime, nonlinear relaxation ($\mathcal{N}$) balances linear amplification ($\mathcal{L}$), effectively decorrelating fluctuation memory. Because the turbulence remains near a stable statistical attractor, the stress--strain relationship is universally structured and locally prescribable. ML excels in this disorganized but statistically well-behaved regime, where data-driven RANS surrogates can reliably interpolate phase space and transfer robustly across geometries.

\noindent \textbf{Regime III: Organized Simplicity (Low $Re$, High $SK/\epsilon$):} In these unsteady deterministic flows, external forcing drives modal bifurcations or strong non-normal transient growth \cite{Schmid2001}, causing fluctuations to acquire non-Markovian memory and coherent organization. Because phase-space dimensionality remains low, direct scale resolution (DNS/LES) is computationally feasible and often preferable to statistical surrogacy. Crucially, this regime is ideal for physics-informed neural networks (PINNs): unlike in broadband turbulence, where PINNs encounter non-uniqueness and spectral stiffness, narrow-band deterministic dynamics tightly constrain the optimization landscape. Consequently, PINNs reliably converge to unique physical solutions directly from governing equations and boundary conditions, serving as robust, model-free solvers.  
    
\noindent \textbf{Regime IV: Emergent Complexity (High $Re$, High $SK/\epsilon$):} Far-from-equilibrium turbulence characterized by massive, long-lived coherent structures embedded within a high-$Re$ background. In this structurally irreducible regime, amplification mechanisms heavily dominate stochastic relaxation. Coherent structures and Reynolds stresses act as topology-dependent, co-emergent phenomena, mathematically precluding \textit{a priori} local constitutive relations \cite{Girimaji2024NJP}. Because full resolution is computationally prohibitive, this regime strictly demands hybrid scale-resolving simulations (SRS) \cite{Spalart1997,Girimaji2006} designed to explicitly compute the organized coherent structures while relying on local models only for the disorganized stochastic background.

By explicitly distinguishing the disorganized stochastic regime (Regime II) from structurally emergent dynamics (Regime IV), this taxonomy demystifies the generalization crisis in data-driven turbulence modeling. It demonstrates that ML-RANS failures in complex flows are not caused by algorithmic limitations or insufficient training data, but by the fundamental misapplication of statistical tools to structurally emergent regimes where non-Markovian memory effects physically invalidate local closure.

\subsection{The Flow-Dependent Nature of Emergent Memory}

A mathematical consequence of the Mori--Zwanzig formulation is that the memory kernel is an emergent operator, not a universal constitutive property. Just as reduced-order representations (like POD/DMD modes) are bespoke to their generating flows, the exact memory kernel is fundamentally tied to the macroscopic flow configuration:
\begin{equation}
K_{ij} = K_{ij}[\mathcal{F}; t, s],
\label{eq:kernel_flow_dep}
\end{equation}
where $\mathcal{F}$ encompasses geometry, boundary conditions, Reynolds number, and forcing mechanisms. Because different flows exhibit distinct global instabilities, their memory kernels are structurally unique.

This establishes a firm theoretical boundary on purely data-driven closures. An ML model optimized over a finite database constructs an approximation $K_\theta = \mathcal{ML}(\mathcal{D}_{\rm train})$. While this surrogate may interpolate within the sampled manifold, it cannot mathematically guarantee extrapolation to unseen instability mechanisms. The challenge in Regime IV is not insufficient data, but the impossibility of learning a universal mapping for a highly flow-specific, emergent quantity. Consequently, the most robust strategy is not to model this history dependence, but to dynamically resolve the memory-bearing coherent structures via scale-resolving simulations (SRS).

\section{Statistical Learning in Emergent Regimes}

Because the memory kernel in Regime IV depends on $\mathcal{F}$, the mapping from local kinematics ($\boldsymbol{\mu}$) to Reynolds stresses ($\mathbf{y}$) is fundamentally \emph{non-injective}. Two distinct geometries ($\mathcal{F}_1, \mathcal{F}_2$) can exhibit identical local kinematics while yielding entirely different stresses ($\mathbf{y}_1 \neq \mathbf{y}_2$). The physically correct mapping relies on global context, governed by $P(\mathbf{y} \mid \boldsymbol{\mu}, \mathcal{F})$. Standard data-driven closures mistakenly approximate this as $P(\mathbf{y} \mid \boldsymbol{\mu})$---an omission justifiable in Regime II where nonlinear relaxation erases memory, but fatal in Regime IV.

A common workaround is to pursue a ``unified'' closure by aggregating training data across diverse geometries $\{\mathcal{F}_k\}_{k=1}^N$. However, omitting global context reduces the problem to a flawed statistical compromise. According to statistical learning theory \cite{CoverThomas2006, Hastie2009}, the optimal predictor $f^*$ that minimizes the mean squared error across this aggregated dataset is the conditional expectation:
\begin{equation}
f^*(\boldsymbol{\mu}) = \sum_{k=1}^N P(\mathcal{F}_k \mid \boldsymbol{\mu}) \, \mathbb{E}[\mathbf{y} \mid \boldsymbol{\mu}, \mathcal{F}_k].
\label{eq:conditional_expectation}
\end{equation}
Physically, $\mathbb{E}[\mathbf{y} \mid \boldsymbol{\mu}, \mathcal{F}_k]$ is the true Reynolds stress, $\mathbf{y}_k$, for geometry $k$. The term $P(\mathcal{F}_k \mid \boldsymbol{\mu})$ acts as a statistical weight, $w_k$, reflecting how frequently the local state $\boldsymbol{\mu}$ appears in geometry $k$ during training. The optimal prediction therefore reduces to a weighted average:
\begin{equation}
f^*(\boldsymbol{\mu}) = \sum_{k=1}^N w_k \mathbf{y}_k.
\end{equation}

For structurally non-injective flows, this mathematical compromise produces a smeared stress tensor that satisfies no valid governing equations. The resulting Bayes error is irreducible. Increasing training data diversity merely incorporates more incompatible mappings into the mixture, exacerbating the error. Thus, poor generalization in Regime IV stems from the fundamental absence of a well-defined functional relationship, not limited model capacity.

In out-of-distribution applications ($\mathcal{F}_{\mathrm{test}}$), these flaws are severely magnified. Because standard neural networks learn statistical correlations rather than invariant causal relationships \cite{Scholkopf2021, Pearl2009}, they lack the physical mechanisms to extrapolate across global context changes, yielding uncontrolled errors.

This clarifies the operational boundaries of data-driven turbulence modeling. The decisive question is not whether AI can emulate the Navier--Stokes equations, but whether a given turbulent state physically admits a coarse-grained statistical description. The turbulence community must pivot toward adaptive, flow-aware computation: deploying ML where the physics admits reliable local modeling, and explicitly resolving the scales where it does not.

\section*{Outlook}
Achieving optimal outcomes in data-driven turbulence modeling requires pivoting from universal, one-size-fits-all closures toward situation-aware AI/ML deployment. Conceptually, this perspective advances the field by grounding ML learnability in complex-systems theory and non-equilibrium physics, demystifying why data-driven surrogates fail in memory-bearing regimes. Practically, it proposes a physics-based recipe for flow-aware simulation parameterized by $SK/\epsilon$, enabling researchers to avoid misapplying ML-RANS while optimizing computational expenditure in scale-resolving simulations. Crucially, this qualitative framework sets the stage for establishing precise quantitative threshold values of $SK/\epsilon$. Preliminary work \cite{Girimaji2026} leverages non-equilibrium thermodynamics \cite{Prigogine1967} toward this goal, laying the groundwork for an \textit{AI adjudicator} capable of evaluating flow states \textit{in situ} to dynamically select physically valid, computationally optimal modeling strategies.

\section*{Conflict of Interest Statement}
None declared.


\end{document}